\documentclass[a4paper,pra,preprint]{revtex4}
\usepackage{hyperref}
\usepackage{amssymb, amsmath}
\usepackage{amsfonts}
\usepackage{comment}
\usepackage{subfigure}
\usepackage{epsfig}
\usepackage{graphics}
\usepackage{anysize}
\usepackage{appendix}
\usepackage[all]{xy}
\usepackage{bm}
\usepackage{pgf,pgfarrows,pgfnodes,pgfautomata,pgfheaps,pgfshade}
\usepackage{graphicx}
\usepackage{bbm}
\usepackage{tikz} 
\usepackage{pgf}

\usetikzlibrary{arrows,decorations.pathmorphing,backgrounds,positioning,fit,petri}
\usetikzlibrary{shapes.multipart}
\usetikzlibrary{shapes.geometric}

\begin{document}

\title{Avoiding Fake State or Bright Light attack on the Single Photon Detector}

\author{Norshamsuri Bin Ali \\
}
\vfill

\date{October 10, 2011}

\maketitle
\section{Fake State detector simulation}\label{Sec:Intro}
		The eavesdropper technique nowadays is already improved from the theoretical perspective to the experimental perspective. The technique now more focusing on the loop holes of the components used such as modulator, laser and detector. These all components actually are classical component which normally being used in the communication system. The technique called "blinding detector" introduce by Vadim Makarov et. al.\cite{V1,Lars1,Lars2,Sebas1,Wiechers} exploit the unavailability of true single photon detector. The detector behavior which is avalanche photo diode (APD) is being used in almost all quantum system and being exploit it vulnerability towards the quantum attacks which is limit the potential to detect the presence of the eavesdropper attack. In this report, I will try to show the loopholes on detector that utilized by the eavesdropper to attack using blinding detector technique which start from the passive quenching APD, active quenching, and gated APD. Secondly, I try to match the countermeasure technique which being proposed by my colleague and I call it "The Counter Measure".

\section{The Blinding detector technique}\label{Blind}
The exploitation of the detector on the quantum system has been exploited since 2005 \cite{V2}and then the most sophisticated and guaranteed that eavesdropper has full control which call blinding detector was introduce in 2007\cite{Sebas1}. The technique is actually to force the detector to its linear region where the eavesdropper attack is unnoticed by the QKD system. The bright illumination light is used to force the detector to this linear region.

\subsection{The passively quenched APD}\label{Sec:passive}
\begin{figure}[ht]
\includegraphics[width=.85\linewidth]{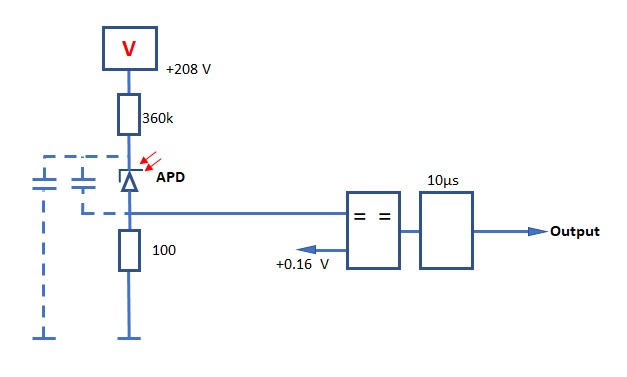}
\caption{\label{fig1}Equivalent circuit diagram for Si SPAPD \cite{V1}.}
\end{figure}

The passively quenched single photon detector is actually the most simple single photon detector \cite{Cova1} and this type of detector only working for the silicon type APD which normally used in the free space QKD system with wavelength range from 500nm-900nm\cite{V1}. The APD circuit as shown in Figure \ref{fig1} shows the equivalent for this detector. As normal APD the biased voltage is used to bias the detector above the breakdown voltage. This is done by the charging the capacitance as shown in Figure \ref{fig1} to the voltage of the biased voltage which in the particular detector by \cite{V1} is about 6-10V and it takes about 1us. This is actually the loopholes that being exploited by bright illumination signal. Since the single photon APD working on the above breakdown voltage it need to be power up by the biased voltage. The single photon APD is discharge the voltage in the capacitance when it detects single photon. This voltage is used as current signal to trigger the next electronic components which is normally comparator. The time taken to power up this capacitor APD's is about $1\mu s$ as report in \cite{V1}.  During the power up, the APD does not work as single photon APD and it is blind and cannot detect any incoming light. This loophole is being used by eavesdropper to always send the incoming signal which will maintain the capacitance voltage is always discharge and let the detector to be blind.

\begin{figure}[ht]
 \includegraphics[width=.85\linewidth]{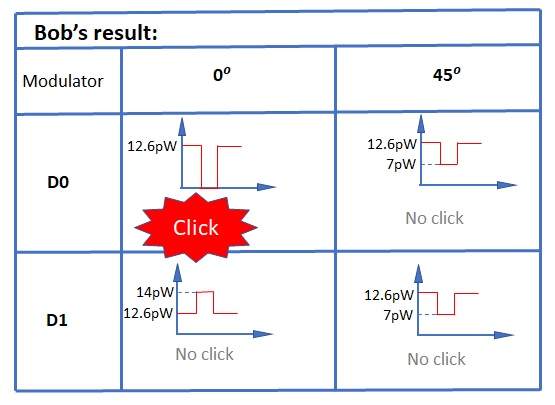}
\caption{\label{fig2}Eavesdropper register click at D0\cite{V1}.}
\end{figure}

 When the eavesdropper need a click \cite{Cova2}  to the detector they just stop the incoming light which about $2\mu s$ as report in \cite{V1} and turning back the bright illumination light. The click will register on the rising edge of the blanking signal to the bright illumination which shows in Figure \ref{fig2}. In this Figure the eavesdropper polarized the light which about $2\mu s$ so that if the Bob measure with the same basis as eve the click will register as wanted by eavesdropper.

\subsection{The active quenched detector}
\begin{figure}[ht]
 \includegraphics[width=.85\linewidth]{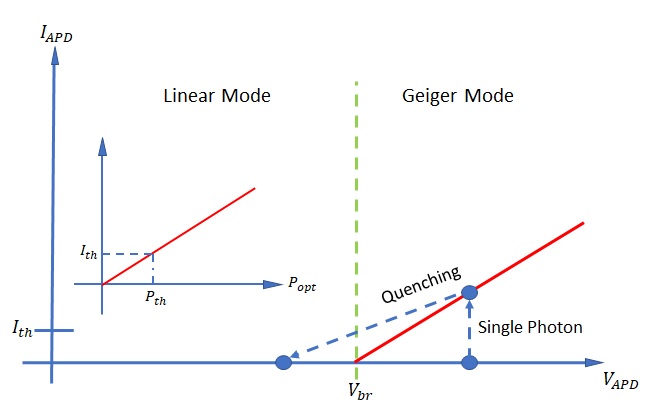}
\caption{\label{fig3}The detector current toward optical power and APD voltage. (a) Linear mode (b) Geiger mode\cite{Lars1}.}
\end{figure}

The active quenched detector mostly based on the \cite{Cova2} which biased the detector above the breakdown voltage to achieved the Gieger mode region as shown in Figure \ref{fig3}. In this region the APD is sensitive to the single photon detection as claim in \cite{Cova1} which gives the rise of performance more than 50\% compare to the passively quenched detector. The active quenched detector is normally given to the Si APD since the biased voltage above the breakdown voltage is always apply (fixed) since the dark count rate is low in this detector type\cite{Cova2}. The same technique is also apply for the InGaAs APD which normally used for the fiber based system but the biased is gated mode to reduce the noises available for this detector type\cite{Cova2}.

\begin{figure}[ht]
\includegraphics[width=.85\linewidth]{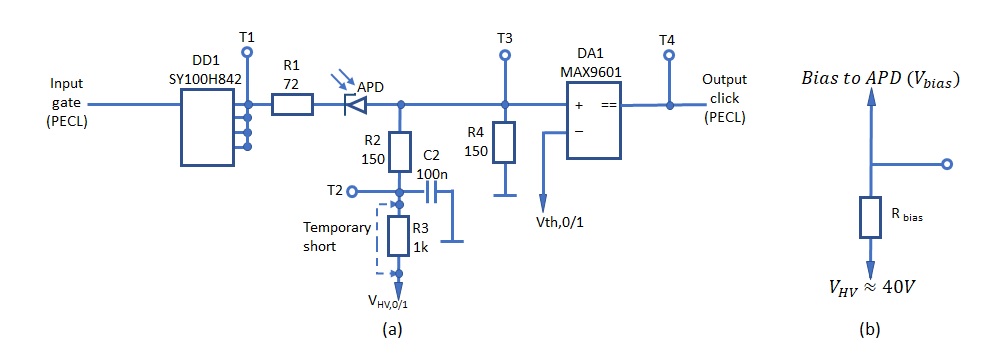}
\caption{\label{fig4}The quenching circuit Si APD (a) complicated \cite{Lars2} (b) simplified \cite{Lars1}.}
\end{figure}

The circuit design is claim in the \cite{Lars2} shown in the Figure \ref{fig4} below is show one of the loopholes that used by eavesdropper to attack the actively quenched detector seen in the Figure \ref{fig4}(a) where the bright illumination is used to attack the APD so that APD fall into the linear region.

The technique actually sending continuous wave light or the pulse bright light to the detector with the frequency starting from $> 70 KHz$ which actually blinding the detector in between the pulses or at every vacuum pulses. This bright light illumination putting more current into the $Rbias$ and increased the voltage drop at the Rbias then actually reduces the bias voltage at T1 as shown in Figure \ref{fig4}(b). These pulses actually bring the bias voltage below the breakdown voltage. Thus, the detector now works in the linear region where every power is correspond to the current as shown in Figure \ref{fig3}(a). Any signal above the $> Pth$ the detector will register the click. Unfortunately the method of attack does not work on \cite{SPCM} but the next attack on the active quenched detector is working which are thermal blinding and DC/DC converter overload.
The thermal blinding is happen when the pulse repetition of the bright illuminating pulses increase up to 1MHz where the TEC current saturated and the temperature increased significantly given to the breakdown voltage rises which bring the fixed bias voltage fall below breakdown voltage. Thus, the detector fall in linear region and the attack as mention previously is working by polarized the light above the $> Pth$ and the detector will click according Eve choice which regard Bob and Eve share the same measurement bases.

\subsection{The gated detector}
As mention above the gated detector is actually actively quenched detector but the material is different which is using InGaAs APD detector and it is normally used in telecom wavelength system. Since the APD biased does not apply directly to the detector but only apply in the gated mode the detector actually biased below breakdown voltage. Since the gated mode signal is actually used to bias the detector above the breakdown voltage and the detector will be in the Geiger mode region during the detection period \cite{Lars1}. In order to keep the detector always in linear mode even during the gated time, the continuous illumination signal is sending where the same phenomena is happen as mention above with the $Rbias$ is used as the loopholes to reduce biased voltage as shown in Figure \ref{fig4}. Then the same attack is applied as in mention in the active quenched detector. The only difference on this attack is to register the click which normally gated by the bias voltage to the detector but based on \cite{Wiechers} the attack is done on the after gate which eve have approximately $10ns$ to attack during this time. Eve cannot attack during the presence of bias voltage since it will bring the APD into the Geiger mode.

\section{The Counter Measure.}
\begin{figure}[ht]
\includegraphics[width=.85\linewidth]{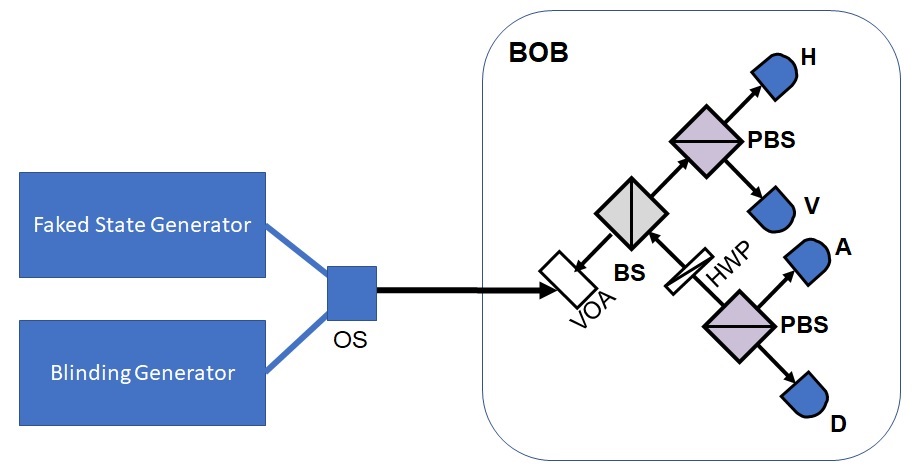}
\caption{\label{fig5} The counter measure setup.The Alice setup is normal setup and between Alice and Bob where Fake State Generator and Blinding Generator being placed. Eve will measure the result from Alice as in intercept resend attack and using these two generators to trigger the click based on the Eve result's. The OS is optical switch, VOA is variable optical attenuator, BS is beam splitter, HWP is half wave plate, and PBS is polarization beam splitter. }
\end{figure}

As for the counter measure the design purposed as shown in Figure \ref{fig5} below where the variable attenuator (VOA) is placed in detector side with only be known by the Bob. This security measure is used to trace any eavesdropper attack in the system. Several reply has been published to counter measure the technique of the detector blinding \cite{Luan,Yuan} but the counter measure that purposed is just to overcome some of the issue on the loopholes in the APD circuitry by no counter measure that can proven on the security proof apply\cite{Lars3,Lars4}. The technique propose is one of the counter measure experimentally viable but still has the loopholes on the other technique such as in the Trojan attack \cite{Ribordy}. I will try to explain this counter measure in the experimental point of view which actually the security measurement should be done in post processing analysis where the presence of eavesdropper can be notified. I divided this counter measure into two sections since the attack on the active quenched detector and the gated detector is fall in the same category and the passively quenched detector fall in the other category.

\subsection{Active quenched detector and the gated detector.}
In the active quenched detector as we already go through also in the gated mode detector, the loopholes exploited are the $R bias$ where the voltage drop in this resistor coursed the biased voltage drop and other technique used to drop the bias voltage by heating the detectors. Then the attack as mention in \cite{Lars1,Lars2,Sebas1,Wiechers} as shown in the Equation \eqref{eq1} and Equation\eqref{eq2} below are the scenario of the technique which to force the detector registers clicks. This limitation has to be followed since detector has fallen to the linear mode as shown in Figure \ref{fig3}(a) where the output power proportionally to the current with should surpass the threshold. In the event of the counter measure, detector indeed fall in the blinded mode but the clicks with the limitation shown in Equation \eqref{eq1} and Equation\eqref{eq2} below will be able to trace the eavesdropper present which mentions by \cite{Lucamarini} that eavesdropper cannot maintain the originality of detection clicks. As mention in the \cite{Lucamarini} eavesdropper can hide is present by reproduce back the P0 and P1 but it will fails to hide the existence if the four detector is monitored. In addition, eve lost all the availability to make sure the detector click according to the state that eavesdropper want since the VOA actually prevent it from happen and eve does not have knowledge on the random sequence that apply Bob in all his measurement either power and polarization.

\begin{align}\label{eq1}
max_i\{ P_{D0;0\%}\}<2\biggl(min_i P_{0\%,i } \biggr)
\end{align}

\begin{align}\label{eq2}
\Theta_t= \frac{min\{ P_{D0;0\%}(t), P_{D1;0\%}(t) \}}{max\{ P_{D0;100\%}(t), P_{D1;100\%}(t) \} }> 0.5
\end{align}

\subsection{Passive quenched detector}
In the passive quenched detector as mention above, the eavesdropper send the bright light illumination which enables Eve to fall the detector in its blinded mode where the recharging of the voltage at the capacitance requires time for the detector to achieve above the breakdown voltage. This time as claim in \cite{V1} is 1us which enable Eve to blind the detector by continuously sending the light before the recharging time so that the capacitance is always discharge and the detector is blind. In order to register the click Eve just vacuum the signal that she wants the detector to click and the detector will click accordingly. This method actually is a bit tricky to counter measure since the vacuum pulses is the event that detector will clicks which actually opposed to the Equation \eqref{eq1}  and Equation\eqref{eq2}  above. The only presence is the detector click which actually will be disturbing by the presence of the VOA. This VOA will caused irregularity in detector clicks which can detect the presence of eavesdropper. The other technique which place on the intensity modulator and the VOA electronic control which the VOA can drop the signal up to $60dB$\cite{OZ} and the intensity modulator can drop the signal up to $20dB$. This actually will enable Bob to drop the signal almost up to $80dB$ and open the signal up to $0dB$ which actually will causes the problematic signal to bright illumination signal to blinding the detector. Bob also can apply the frequency method in the attenuation together with the polarization in order to scan the presence of the eavesdropper. This sure will cause problem to eavesdropper since the total opening of the attenuation will kill detector if eavesdropper try to counter the power loss of the attenuation of $80dB$.

\end{document}